\documentclass{iau}

\usepackage{amsmath}
\usepackage{graphicx}
\usepackage{multirow}
\usepackage{xspace}

\newcommand\aap{A\&A}                
\newcommand\aj{AJ}                   
\newcommand\apj{ApJ}                 
\newcommand\apjl{ApJ}                
\newcommand\mnras{MNRAS}             
\newcommand\nat{Nature}              

\newcommand{\ionic}[1]{$\,${\sc #1}}

\newcommand{\forb}[2]{[{#1}\ionic{#2}]}
\newcommand{\forbl}[3]{\forb{#1}{#2}\xspace$\lambda$#3\xspace}
\newcommand{\forbr}[4]{\forb{#1}{#2}\xspace$\lambda$#3/$\lambda$#4\xspace}

\newcommand{\perm}[2]{#1\ionic{#2}}

\newcommand{\hi}{\perm{H}{i}\xspace}

\newcommand{\hb}{H$\beta$\xspace}

\begin{document}

\lefttitle{C. Morisset et al.}
\righttitle{Abundance determination in PNe}

\journaltitle{Planetary Nebulae: a Universal Toolbox in the Era of Precision Astrophysics}
\jnlDoiYr{2023}
\doival{10.1017/xxxxx}
\volno{384}

\aopheadtitle{Proceedings IAU Symposium}
\editors{O. De Marco, A. Zijlstra, R. Szczerba, eds.}
 
\title{Abundance determination in PNe: How to deal with large chemical inhomogeneities
}

\author{C. Morisset$^{1,2}$, J. Garcia-Rojas$^{3,4}$, V. Gomez-Llanos$^{3,4}$ \& H. Monteiro$^5$}
\affiliation{
$^{1}$Instituto de Astronom\'ia (IA), Universidad Nacional Aut\'onoma de M\'exico, Apdo. postal 106, C.P. 22800 Ensenada, Baja California, M\'exico\\
$^2$ Instituto de Ciencias Físicas, Universidad Nacional Autónoma de México, Av. Universidad s/n, 62210 Cuernavaca, Mor., México\\
$^{3}$Instituto de Astrof\'isica de Canarias, E-38205 La Laguna, Tenerife, Spain\\
$^{4}$Departamento de Astrof\'isica, Universidad de La Laguna, E-38206 La Laguna, Tenerife, Spain\\
$^{5}$Instituto de F\'{\i}sica e Qu\'{\i}mica, Universidade Federal de Itajub\'a, Av. BPS 1303-Pinheirinho, 37500-903, Itajub\'a, Brazil 
}

\begin{abstract}
Abundance determinations in planetary nebulae (PNe) are crucial for understanding stellar evolution and the chemical evolution of the host galaxy.

We discuss the complications involved when the presence of a metal-rich phase is suspected in the nebula. We demonstrate that the presence of a cold region emitting mainly metal recombination lines necessitates a detailed treatment to obtain an accurate assessment of the enrichment of this cold gas phase. 


\end{abstract}


\maketitle

\section{Introduction}

For more than 80 years, the abundance discrepancy (AD) has been a challenge in the field of abundance determination in ionized nebulae. This discrepancy is the fact that the ionic abundances relative to hydrogen (X$^{+i}$/H$^+$) obtained from observations of weak recombination lines (RLs) of ions heavier than helium are systematically higher than those obtained from much brighter collisionally excited lines (CELs) corresponding to the same ion \citep{1942Wyse_apj95}.  This discrepancy is generally parameterized through the abundance discrepancy factor \citep[ADF,][]{2003Tsamis_mnra345}, which is the ratio between RLs and CELs abundances determined for a given ion (i.~e. ADF = [X$^{+i}$/H$^+$]$_{RLs}$/[X$^{+i}$/H$^+$]$_{CELs}$). Recently, it has been found that a scenario in which the presence of temperature fluctuations \citep{1967Peimbert_apj150} in the zones where the gas is most ionized, possibly produced due to the stellar feedback effect, may be responsible for the AD in H~{\sc ii} regions \citep{2023Mendez-Delgado_Nat618}. However, this scenario does not appear to work in the case of planetary nebulae \citep[PNe;][these proceedings]{2023Mendez-Delgado_arXiv231110280M}.

The problem is further complicated when considering that in some PNe, the AD is much greater (ADF $\geq$ 10) than that measured in the vast majority of ionized nebulae (ADF $\sim$ 2--4). In these cases, a binary star that has gone through a common-envelope phase is usually present. It has been shown that a large part of the AD is caused by two plasma components with very different physical conditions. The first component has temperatures similar to those of typical photoionized nebulae with a spectrum mainly composed of RLs of H and He and CELs of heavy elements. The second component is much colder and emits almost no CELs, but its spectrum of RLs of heavy elements is very strong \citep{2015Corradi_apj803,2018Wesson_mnra480}. The presence of this cold component significantly affects the determination of the physical conditions of the gas and the determination of the chemical abundances in both components \citep{2020Gomez-Llanos_mnra498,2022Garcia-Rojas_mnra510}.

\citet{2020Gomez-Llanos_mnra497} developed a grid of photoionization models that combine cold metal-rich clumps embedded in a normal metallicity region and found that the real contrast between chemical abundances relative to H in both components (the Abundance Contrast Factor, ACF) could be different from the observationally computed ADF. These results call for a review of the methodology for computing ion abundances in this type of objects.

\section{Using a correct temperature for \hb}\label{sec:TeHb}

In the following, we use the same formalism and equations as in Gómez-Llanos et al. (2024, in prep.).

The classical way of determining ionic abundances from emission line is using the following relations:

\begin{equation}
    I_\lambda = \int_V n_{\rm e} . n_{X^i} . \epsilon_\lambda(T_{\rm e}, n_{\rm e}) dv
    \simeq \frac{X^i}{H^+} . n_{\rm e}^2 . \epsilon_\lambda(T_{\rm e}, n_{\rm e}) . V,
    \label{eq:ion_ab_classical1}
\end{equation}

where $n_{\rm e}$ and $n_{X^i}$ are, respectively, the densities of electrons and of the ions under consideration, ${\epsilon_\lambda(T_{\rm e}, n_{\rm e})}$ is the line emissivity in erg/s.cm$^3$, $X^i / H^+$ is the ionic abundance to be determined, $T_{\rm e}$ is the electron temperature (but see below).  The integral is carried out over the volume $V$ of the nebula. In the case of \hb line this relation is:
\begin{equation}
    I_\beta = \int_V n_{\rm e} . n_{H^+} . \epsilon_\beta(T_{\rm e}, n_{\rm e}) dv
    \simeq n_{\rm e}^2 . \epsilon_\beta(T_{\rm e}, n_{\rm e}) . V,
    \label{eq:ion_ab_classicalHb}
\end{equation}

neglecting the contribution of elements other than H to free electrons.  Combining both relations leads to: 
\begin{equation}
    \frac{X^i}{H^+} = \frac{I_\lambda }{I_\beta} . \frac{\epsilon_\beta(T_{\rm e}, n_{\rm e})}{\epsilon_\lambda(T_{\rm e}, n_{\rm e})},
    \label{eq:ion_ab_classical}
\end{equation}

where ${I_\lambda }/{I_\beta}$ is the line ratio involving the $X^i$ ion. 
Here, we find the first potential problem: the temperature and density used in both emissivities are not necessarily the same. It is very common in detailed analysis to consider two or even tree zones in the considered nebula, the difference between the zones being described by the main ions it contains. 
For example, a two-zone description could have a low ionization zone where the temperature is determined from \forbr{N}{ii}{5755}{6584}, while the high ionization zone is described by the \forbr{O}{iii}{4363}{5007} temperature diagnostic. The important issue here is the temperature (and density in a less important way) used in the \hb emissivity: as the \hb line is the same in both zones, its emissivity should not depend on where the metal line is coming from. The emissivity of the \hb line must be obtained from a pair of $T_{\rm e}, N_{\rm e}$ representative of the whole $H^+$ region, which encloses both zones.

As an illustration, we consider a nebula where the low ionization zone has a temperature of 12,000\,K, while the high ionization zone is at 9,000\,K. The \hi line can be considered at 10,000\,K. The classical way to determine abundances, using the same electron temperature for the metal line and for \hb  would underestimate the $N^+/H^+$ ionic abundance in the low ionization zone (using \forbl{N}{ii}{6584}) by 15\%, while it would overpredict the $O^{++}/H^+$ ionic abundance in the high ionization zone (using \forbl{O}{iii}{5007}) by 9\%.

The PyNeb \verb|getIonabunce| method gives the user the opportunity to specify the electron temperature for the line under consideration, as well as the temperature for the \hb line used to normalize the observation: 
\begin{verbatim} 
O3.getIonAbundance(int_ratio=350, tem=9000, den=100., wave=5007, 
                   Hbeta=100, tem_HI=10000)
\end{verbatim}


\section{Cold regions disguised as hot ones} \label{sec:cold}

In the case of a metal-rich cold phase, the dominant channel to produce emission lines such as \forbl{N}{ii}{5755} or \forbl{O}{iii}{4363} is not longer the radiative decay following a collisional excitation of N$^+$ or O$^{++}$ ions, but rather the recombination of N$^{++}$ to N$^{+}$ or O$^{3+}$ to O$^{++}$, respectively.

This phenomenon has already been reported in observations \citep[e.g.][]{2016Jones_mnra455, 2016Garcia-Rojas_apjl824, 2022Garcia-Rojas_mnra510} and in theoretical studies \citep[e.g.][]{2020Gomez-Llanos_mnra497}. Failing to consider the contribution of the recombination to the auroral lines leads to significantly inaccurate estimations of the electron temperature and subsequently the chemical abundances.

\section{Detailed abundance determinations for two-components plasma}

In the case of two different phases of gas mixed in the line of sight, the previous relations lead to the following.

\begin{equation}
\begin{split}
I_\lambda = & I_\lambda^w + I_\lambda^c  \\
          = & [\frac{X^i}{H^+}]^w \cdot (n_{\rm e}^w)^2 \cdot \epsilon_\lambda(T_{\rm e}^w, n_{\rm e}^w) \cdot V^w  + \\
            & [\frac{X^i}{H^+}]^c \cdot (n_{\rm e}^c)^2 \cdot \epsilon_\lambda(T_{\rm e}^c, n_{\rm e}^c) \cdot V^c \\
          = & [\frac{X^i}{H^+}]^w \cdot [N_{\rm e}.n_{\rm e}]^w \cdot \epsilon_\lambda(T_{\rm e}^w, n_{\rm e}^w) + \\
            & [\frac{X^i}{H^+}]^c  \cdot [N_{\rm e}.n_{\rm e}]^c \cdot \epsilon_\lambda(T_{\rm e}^c, n_{\rm e}^c),
\end{split}
\end{equation}
 the case of \hb being:
\begin{equation}
\begin{split}
I_\beta = & I_\beta^w +  I_\beta^c  \\ 
        = & (n_{\rm e}^w)^2 \cdot \epsilon_\beta(T_{\rm e}^w, n_{\rm e}^w). V^w + (n_{\rm e}^c)^2 \cdot \epsilon_\beta(T_{\rm e}^c, n_{\rm e}^c). V^c \\
        = & [N_{\rm e}.n_{\rm e}]^w \cdot \epsilon_\beta(T_{\rm e}^w, n_{\rm e}^w) + [N_{\rm e}.n_{\rm e}]^c \cdot \epsilon_\beta(T_{\rm e}^c, n_{\rm e}^c) \\
        = & (1-\omega) \cdot I_\beta + \omega \cdot I_\beta,
\end{split}
\end{equation}

where the $w$ and $c$ superscripts are used for the warm and cold regions, respectively, and $\omega$ is defined as the weight of the cold region in the $I_\beta$ emission: $\omega = I_\beta^c/I_\beta$. The total number of electrons in each region is $N_{\rm e} = n_{\rm e} . V$.

Combining the 2 previous equations, we obtain:

\begin{equation}
\begin{split}
\frac{I_\lambda}{I_\beta} = & [\frac{X^i}{H^+}]^w \cdot (1-\omega) \cdot \frac{\epsilon_\lambda(T_{\rm e}^w, n_{\rm e}^w)}{\epsilon_\beta(T_{\rm e}^w, n_{\rm e}^w)} + \\
                            & [\frac{X^i}{H^+}]^c  \cdot \omega \cdot \frac{\epsilon_\lambda(T_{\rm e}^c, n_{\rm e}^c)}{\epsilon_\beta(T_{\rm e}^c, n_{\rm e}^c)}.
\end{split}
\label{eq:ion_ab_2comp}
\end{equation}

When a line is mainly emitted through collisional excitation followed by radiative transitions, the ratio of its emissivities $\epsilon_\lambda/\epsilon_\beta$ vanishes at low electron temperature, and the line is produced exclusively by the warm region. On the other hand, when a line originates from recombination followed by radiative transitions, the ratio of emissivities $\epsilon_\lambda/\epsilon_\beta$ remains almost independent of the electron temperature. The contribution of the warm (cold) region to the observed line intensity depends primarily on $1-\omega$ ($\omega$ respectively), as well as on the ionic abundance of the emitting ion.

Using the findings of bi-abundance models built by \citet{2020Gomez-Llanos_mnra497} and depicted in their Fig.~18, in the following we explore the hypothesis that \forb{O}{iii} and \perm{O}{ii} come from exclusive regions (warm and cold, respectively). The ionic abundances for each region are determined from different lines using the following equations:

\begin{equation}
    [\frac{X^i}{H^+}]^w = \frac{I_\lambda }{(1-\omega)\cdot I_\beta} \cdot \frac{\epsilon_\beta(T_{\rm e}^w, n_{\rm e}^w)}{\epsilon_\lambda(T_{\rm e}^w, n_{\rm e}^w)}, 
    \label{eq:ion_ab_w}
\end{equation}

when $I_\lambda^c = 0$, and:
\begin{equation}
    [\frac{X^i}{H^+}]^c = \frac{I_\lambda }{\omega \cdot I_\beta} \cdot \frac{\epsilon_\beta(T_{\rm e}^c, n_{\rm e}^c)}{\epsilon_\lambda(T_{\rm e}^c, n_{\rm e}^c)},
    \label{eq:ion_ab_c}
\end{equation}

when $I_\lambda^w = 0$.

It is important to note here that the $T_{\rm e}$ and $n_{\rm e}$ values used to compute the \hb emissivity should correspond to the specific region under consideration (warm or cold). Refer to Sec.~\ref{sec:TeHb} for the case of the warm region.

The ionic abundances derived from both collisionally excited lines and recombination lines using the generic eq.~\ref{eq:ion_ab_classical} are both underestimated by a factor close to $1/(1-\omega)$ and $1/\omega$ respectively. If the effect of $\omega$ is not taken into account, the ratio between the two determinations is the classical ADF. Thus, ACF is linked to ADF by a factor of $(1-\omega)/\omega$, which can be notable. 
Hence, there is a clear necessity to determine $\omega$ for each spaxel in the images to attain reliable abundances.

The determination of the weight of the cold region $\omega$ is a challenging task. It can be computed by considering that the hydrogen emission arises from both regions, and that the Balmer or Paschen jump normalized to an \hi line is indicative of the emission contribution from each region. This ratio is commonly employed to determine the so-called Balmer or Paschen temperature. In the case of two regions with very different temperatures, the derived temperature from the Balmer or Paschen jump may not accurately reflect the actual temperature of any plasma component within the observed object.

The theoretical Paschen jump and H-lines are computed as the sum of contributions from both the cold and the warm regions:

\begin{equation}
\label{eq:pj_omega}
\begin{split}
    PJ\left(\omega, T_{\rm e}^w, T_{\rm e}^c, n_{\rm e}^w, n_{\rm e}^c\right) = & [N_{\rm e}.n_{\rm e}]^w \cdot C_{8100}(T_{\rm e}^w, n_{\rm e}^w) + [N_{\rm e}.n_{\rm e}]^c \cdot C_{8100}(T_{\rm e}^c, n_{\rm e}^c) -\\
                                                & [N_{\rm e}.n_{\rm e}]^w \cdot C_{8400}(T_{\rm e}^w, n_{\rm e}^w) - [N_{\rm e}.n_{\rm e}]^c \cdot C_{8400}(T_{\rm e}^c, n_{\rm e}^c) \\
      = & \left[(1-\omega)\cdot \frac{I_\beta}{\epsilon_\beta(T_{\rm e}^w, n_{\rm e}^w)} \cdot (C_{8100}(T_{\rm e}^w, n_{\rm e}^w) - C_{8400}(T_{\rm e}^w, n_{\rm e}^w))\right] +\\
    &   \left[\omega \cdot \frac{I_\beta}{\epsilon_\beta(T_{\rm e}^c, n_{\rm e}^c)} \cdot (C_{8100}(T_{\rm e}^c, n_{\rm e}^c) - C_{8400}(T_{\rm e}^c, n_{\rm e}^c))\right] 
\end{split}
\end{equation}

\begin{equation}
\begin{split}
    I_{9229}\left(\omega, T_{\rm e}^w, T_{\rm e}^c, n_{\rm e}^w, n_{\rm e}^c\right)  = &(1-\omega) \cdot \frac{I_\beta}{\epsilon_\beta(T_{\rm e}^w, n_{\rm e}^w)} \cdot \epsilon_{9229}(T_{\rm e}^w, n_{\rm e}^w) + \\ 
                                                       & \omega \cdot \frac{I_\beta}{\epsilon_\beta(T_{\rm e}^c, n_{\rm e}^c)} \cdot \epsilon_{9229}(T_{\rm e}^c, n_{\rm e}^c)
\end{split}
\end{equation}
In the first equation, we omit explicitly stating the dependencies of the continuum on He$^{+}$/H$^+$ and He$^{2+}$/H$^+$, but they are considered. 

\begin{figure}
	\centering
	\includegraphics[width=12.cm]{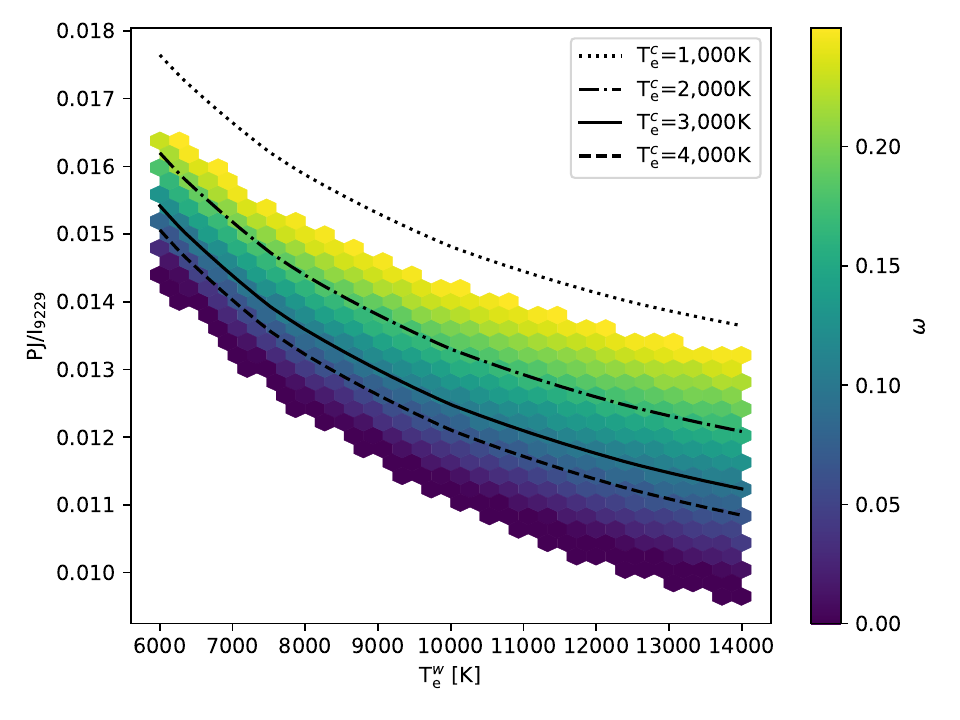}
	\caption{Colored hexagons show the variation of $\omega$ when $T_{\rm e}^w$ varies from 6,000 to 14,000~K and the normalized Paschen jump PJ/I$_{9229}$ varies from 0.010 to 0.017. Fixed values are set to: $T_{\rm e}^c$ = 3,000~K, $n_{\rm e}^w$ = $n_{\rm e}^c$ = 1,000~cm$^{-3}$, He$^{+}$/H$^{+}$ = 0.1, and He$^{2+}$/H$^{+}$ = 0.005. Lines of different styles show the values of  PJ/I$_{9229}$ for different values of $T_{\rm e}^c$ and a fixed value of $\omega$ = 0.1.}\label{fig:omega}
\end{figure}

The normalized Paschen jump PJ/I$_{9229}$ is computed using PyNeb \citep{2015Luridiana_aap573} across a grid of $T_{\rm e}^w$, $T_{\rm e}^c$, and $\omega$, and constant values for $n_{\rm e}^w$, $n_{\rm e}^c$, He$^{+}$/H$^{+}$, and He$^{2+}$/H$^{+}$. These numerical values must be adapted for each object. As an illustration, the values of PJ/I$_{9229}$ obtained for $T_{\rm e}^w$ between 6,000 and 14,000~K, and $\omega$ between 0 and 0.25, and fixed values of $T_{\rm e}^c$ = 3,000~K, $n_{\rm e}^w$ = $n_{\rm e}^c$ = 1,000~cm$^{-3}$, He$^{+}$/H$^{+}$ = 0.1, and He$^{2+}$/H$^{+}$ = 0.005 are used in Fig.~\ref{fig:omega} to draw as colored hexagons the variations of $\omega$ relative to PJ/I$_{9229}$ and $T_{\rm e}^w$. Lines are also showing the effect of changing $T_{\rm e}^c$ to 1,000, 2,000, 3,000 (the same value used for the colored hexagons), and 4,000~ K for the case $\omega$ = 0.1.

A numerical algorithm (which can be an Artificial Neuron Network) is then used to interpolate the inverse problem to predict $\omega$ from T$_{\rm e}^w$, T$_{\rm e}^c$, and PJ/I$_{9229}$.

In the case of NGC~6153, G\'omez-Llanos et al. (2024, in prep.) generate a spatially resolved map of $\omega$ and use it to compute the ACF of O$^{2+}$.  

\section{Conclusions}

As highlighted by \citet{2022Richer_aj164} regarding the PN  NGC\,6153, ``reality is complicated'' when determining the chemical abundances in PNe, particularly in those with high ADFs. It is essential to pay close attention to details when attempting to obtain accurate chemical abundances in these objects, as has been shown in this proceeding. \\

CM acknowledges funding from PAPIIT/UNAM grants IN101220 and IG101223. JG-R and VG-LL acknowledge funding from the Canarian Agency for Research, Innovation and Information Society (ACIISI), of the Canary Islands Government, and the European Regional Development Fund (ERDF), under grant with reference ProID2021010074. JG-R also acknowledges the support of the Spanish Ministry of Science and Innovation (MICINN) through the Spanish State Research Agency, under the Severo Ochoa Centers of Excellence Program 2020-2023 (CEX2019-000920-S).


\input{Morisset.bbl}


\end{document}